\documentclass[conference]{IEEEtran}
\usepackage{blindtext, graphicx}
\ifCLASSINFOpdf
\else
\fi
\usepackage[cmex10]{amsmath}
\usepackage{amssymb,amsthm}
\usepackage{balance}
\begin{document}

\setlength{\abovedisplayskip}{.2cm}
\setlength{\belowdisplayskip}{.2cm}

\title{Noise Sensitivity of Teager-Kaiser Energy \\Operators and Their Ratios}
\author{\IEEEauthorblockN{Pradeep Kr. Banerjee and Nirmal B. Chakrabarti}
\IEEEauthorblockA{Indian Institute of Technology Kharagpur\\
Email: \{pradeep.banerjee, nirmalbc\}@gmail.com}}
\maketitle

\begin{abstract}
The Teager-Kaiser energy operator (TKO) belongs to a class of autocorrelators and their linear combination that can track the instantaneous energy of a nonstationary sinusoidal signal source. TKO-based monocomponent AM-FM demodulation algorithms work under the basic assumption that the operator outputs are always \emph{positive}. In the absence of noise, this is assured for pure sinusoidal inputs and the \emph{instantaneous} property is also guaranteed. Noise invalidates both of these, particularly under small signal conditions. Post-detection filtering and thresholding are of use to reestablish these at the cost of some time to acquire. Key questions are: (a) how many samples must one use and (b) how much noise power at the detector input can one tolerate. Results of study of the role of delay and the limits imposed by additive Gaussian noise are presented along with the computation of the cumulants and probability density functions of the individual quadratic forms and their ratios. 
\end{abstract}

\IEEEpeerreviewmaketitle

\section{Introduction}
There are many applications in communication and control where decisions are made on the basis of the ratios of quantities, measured or computed, which are compared with thresholds. The present work is concerned with situations, where variables are outputs of a linear combination of quadratic forms in correlated Gaussian random variables and their ratios. 
Such quantities arise in a variety of contexts including, for example, monocomponent AM-FM demodulation using a special class of quadratic detectors called the Teager-Kaiser Energy Operator (TKO), maximum ratio diversity combining schemes to combat fading, directivity optimization of antenna arrays, calculations of SINR and various error and outage probabilities, etc. This work primarily examines the noise sensitivity of the TKO by studying the statistical distribution of the operator outputs and their ratios. 

The TKO is an important class of product correlators used for nonstationary signal analysis. Starting in the early 1980s, Herbert M. Teager and Shushan M. Teager [1] developed several tools, including the energy operator in a series of experiments for modeling nonlinear structures in speech signals, caused by modulations and turbulence. Following their pioneering work, the operator was systematically introduced by Kaiser [2] and was shown to track the energy of a simple harmonic oscillator. The past  two decades have witnessed extensive research showing the usefulness of the operators in diverse fields, including AM-FM decomposition for monocomponent signals [3], [4] and aided by the use of filter banks for multicomponent signals [5], speech analysis [4], image texture analysis [8], cochannel and adjacent-channel signal separation [7], frequency and time discrimination [9], [10], etc. to name a few.

For continuous-time signals, $x(t)$, $y(t)$, the kernel $[x,y] \equiv \dot x y - x\dot y$ measures the instantaneous differences in the relative rate of change between $x$ and $y$ [11], where the overdots represent time derivatives. If $x$ is the Hilbert transform of $y$ so that $x$ and $y$ form an analytic pair, $[x,y]$ gives the intensity-weighted phase derivative and the instantaneous frequency (IF) is defined as $\omega(t)=\tfrac{\dot x y - x\dot y}{x^2+y^2}$. 
If $y=\dot x$, then we get the continuous-time TKO: $\Psi [x]\equiv [x,\dot x]=\dot{x}^2-x\ddot{x}$.
For a monocomponent nonstationary sinusoid $x(t)$ with time-varying amplitude $a(t)$ and instantaneous frequency $\omega(t)$, the operator applied to the signal and its derivative yield respectively, $\Psi [x(t)] \approx a^2(t)\omega^2(t)$ and $\Psi [\dot x(t)] \approx a^2(t)\omega^4(t)$, where the approximations hold if the bandwidths of the amplitude/frequency modulating signals are reasonably small compared to the carrier frequency. Under the assumption that the operator outputs are always positive, these observations have motivated the continuous \emph{energy separation algorithm} (ESA) [3], [4] for AM-FM decomposition,
\begin{align*}
\omega^2(t) \approx \frac{\Psi [\dot x(t)]}{\Psi [x(t)]}, \hspace{.6cm} a^2(t) \approx \frac{(\Psi [x(t)])^2}{\Psi [\dot x(t)]} \tag{1}
\end{align*}

The discrete-time version of the operator is derived by replacing $x(t)$ with sequences $x[n]=x(nT)$ and derivatives $\dot x$ with two-sample backward differences $(x[n]-x[n-1])/T$ where $T$ is the sampling interval. The continuous-time operator then reduces (up to one sample shift) to the discrete version given by $\Psi [x[n]]=(x^2[n]-x[n-1]x[n+1])/T^2$ [2]. We introduce a general four-sample symmetric kernel as a linear combination of two correlators, viz., $\Psi_p^q [x[n]]=(x[n-p]x[n+p]-x[n-q]x[n+q])/T^2$. For $p=0$, $q=1$, $\Psi_p^q$ reduces to the conventional discrete-time TKO $\Psi_0^1$ [2]. 
The effect of applying $\Psi_0^q$ at a sampling rate of $f_s=1/T$ is equivalent to applying $\Psi_0^1$ at a sampling rate  of $f_s/q$, so that $q$ defines the underlying sub-rate processing. Replacing the forward and backward sample delays with time delays, and ignoring the scaling factor $T$, $\Psi_p^q$ as a function of the delay parameters $(t_p,t_q)$ becomes 
\begin{align*}
\Psi_p^q(x)=x(t-t_p)x(t+t_p)-x(t-t_q)x(t+t_q),\hspace{1mm} p < q \tag{2}
\end{align*}

A distinct advantage of the TKO is its superior localization property. 
When the input to the operator is the sum of a deterministic signal and a Gaussian noise process with a specified covariance function, the discrete operator kernel is an indefinite quadratic form in nonsingular normal vectors. In the absence of noise, conditions for the positivity of the operator output can be expressed in terms of the concavity of the logarithm of the signal magnitude [6] that hold in general for pure sinusoids and the instantaneous property is also guaranteed. Noise invalidates both of these, particularly under small signal conditions. The frequency computation involves noisy differentiation operations and a short time window that precludes any scope for averaging. Because of the inherent nonlinearity, significant noise$\times$noise and signal$\times$noise perturbations exist. The effect of these terms becomes pronounced when squaring or ratio operations are employed to estimate signal parameters, and a suitable post-filtering action is necessary at the cost of some time to acquire. For the existence of moments of the ratio, the behavior of the density of the denominator near singularities plays a vital role. Under small signal conditions, large amplitudes can occur frequently when the denominator assumes infinitesimally small values, so that the integral defining the moments diverges because of long tails in the density function. A way out is to eliminate negative outputs and outputs near zero by conditioning all measurements of the ratios on the event that the denominator exceeds some detection threshold---the measurement is ignored if threshold detection fails to occur, thus precluding a division by zero [12]. Thresholding retains the speed advantage of the operator and is a viable alternative, when filtering is not an option. 

Earlier work has examined the problem of noise and found expressions for the variance of the amplitude and frequency estimates. Kaiser [2] briefly mentioned results for the discrete operator in zero-mean additive white Gaussian noise. Bovik and Maragos \emph{et al.} [5] gave a more extensive treatment of multiband demodulation and developed the basic statistical properties of the TKO. A spline-based ESA was proposed in [13] for robust estimation in the presence of noise. A more recent work on short-term energy estimation in additive noise is given in [14]. Statistical distributions of the operator outputs and their ratios have not received any attention. This work is primarily aimed at filling this gap by evaluating the cumulants and probability density functions of the individual operator outputs and their ratios. 

\section{Properties of the TKO in Relation to \\Frequency Discrimination}
The symmetric discrete TKO operator kernel implements a translation-invariant, sliding-window filter that integrates energy detection with frequency filtering [15]. The frequency responses of the three-sample input operator $\Psi_0^q$ and four-sample input operator $\Psi_p^q$ have an important difference. 
For a single sine wave of unit amplitude and frequency $\omega$, the frequency response $\Psi_0^q(\omega)=1-\cos(2\omega t_q)$ is unipolar and periodic. At low frequencies ($\omega T < \pi/4q$) when the linear approximation $\sin(\omega T) \approx  \omega T$ holds,  $\Psi_0^q(\omega) \propto \omega^2$. Traditionally, this range is used for AM-FM demodulation. 
For a periodic input signal, it is easy to verify that the power in the odd harmonics can be obtained by averaging the output of a three-sample input TKO $\Psi_0^q$ by adjusting the delay $t_q$ to correspond to a phase difference of $\pi/2$ for the fundamental frequency.

On the other hand, the frequency response $\Psi_p^q(\omega)=\cos(2\omega t_p)-\cos(2\omega t_q)=2\sin\omega(t_q-t_p)(t_q+t_p)$ is bipolar and is not proportional to $\omega^2$ at low frequencies. The response is zero at $\omega=0$ and when either $\omega(t_q-t_p)$ or $\omega(t_q+t_p)$ is a multiple of $\pi$. 
One observes regions where the response is linear around the zeros of $\Psi_p^q(\omega)$. The location of the extrema of the frequency response can be identified as a solution of the equation, $t_p\sin(2\omega t_p)=t_q\sin(2\omega t_q)$. A discriminator characteristic thus results. When the input consists of several sine waves of amplitudes $a_k$ and frequencies $\omega_k$, i.e., $x(t)=\sum a_k\sin(\omega_kt)$, one gets a dc term, $\sum a_k^2[\sin^2(\omega_kt_q)-\sin^2(\omega_kt_p)]$ and a time-varying cross-term, which is a combination of the sum and difference frequency terms arising out of mixing operations. A typical cross-term component is of the form (for $k < l$): ${a_k}{a_l}\{ \cos ({\omega _k} - {\omega _l})t[\cos({\omega _k} + {\omega _l}){t_p} - \cos ({\omega _k} + {\omega _l}){t_q}] - \cos ({\omega _k} + {\omega _l})t[\cos ({\omega _k} - {\omega _l}){t_p} - \cos ({\omega _k} - {\omega _l}){t_q}]\}$.
The choice of a symmetric kernel ensures that all second harmonic components get cancelled. When the input is differentiated to give $\dot x(t)=\sum \omega_ka_k\cos(\omega_kt)$, all the cross-term components are multiplied by $\omega_k\omega_l$. Ordinarily, for monocomponent signals, the dc term obtained from the TKO for the signal and its derivative are used to find estimates of IF and amplitude. For demodulating multicomponent signals, filter banks are used [5]. When two frequencies are present simultaneously, the $\Psi_p^q$ outputs at the sum and difference frequency cross-terms are of equal amplitude. If the ratio of these frequencies is determined, one can utilize this information in combination with the $\Psi_p^q(x)$ and $\Psi_p^q(\dot x)$ outputs to estimate amplitude and frequency parameters. So, if the signals are closely-spaced [9] or belong to the same filter bank, frequency discrimination using the sum and difference frequency cross-terms generated by a suitable combination of correlators is a viable alternative. 

\section{Experiments: TKO Outputs for a Sum of Two Sinusoids and Sinusoid Plus Noise Input}
Given the wide usage of TKOs in IF estimation for monocomponent signals, it is important to assess the effect of noise or of an interfering signal upon the IF estimates, and check for any physical inconsistencies thereof. The situation is best analyzed by considering the effect of a sinusoidal interference. 
Considering the most general form of a sum of two sinusoids with positive amplitudes $a_1$ and $a_2$, incommensurate frequencies $f_1$ and $f_2$, and independently uniformly distributed phases $\varphi_0$ and $\psi_0$,
\begin{align*}
  x(t;{a_1},{f_1},{a_2},{f_2}) &= {a_1}\cos (2\pi {f_1}t + {\varphi _0}) + {a_2}\cos (2\pi {f_2}t + {\psi _0}) \hfill \\
   &= {a_1}\cos \varphi  + {a_2}\cos \psi  \hfill
\end{align*}
Without loss of generality, the model can be simplified by taking into account the amplitude and frequency ratios, $a = a_2/a_1$, $f = f_2/f_1$ and $\theta_0=\psi_0-\varphi_0$ as the following analysis is sensitive only to such relative parameters. The simplified signal model becomes
\begin{align*}
x(t;a,f) = \cos (2\pi t) + a\cos (2\pi ft + {\theta _0}) \tag{3}
\end{align*}
\begin{figure*}[!t]
\centering
\includegraphics[width=6.5in,height=4.0in,keepaspectratio]{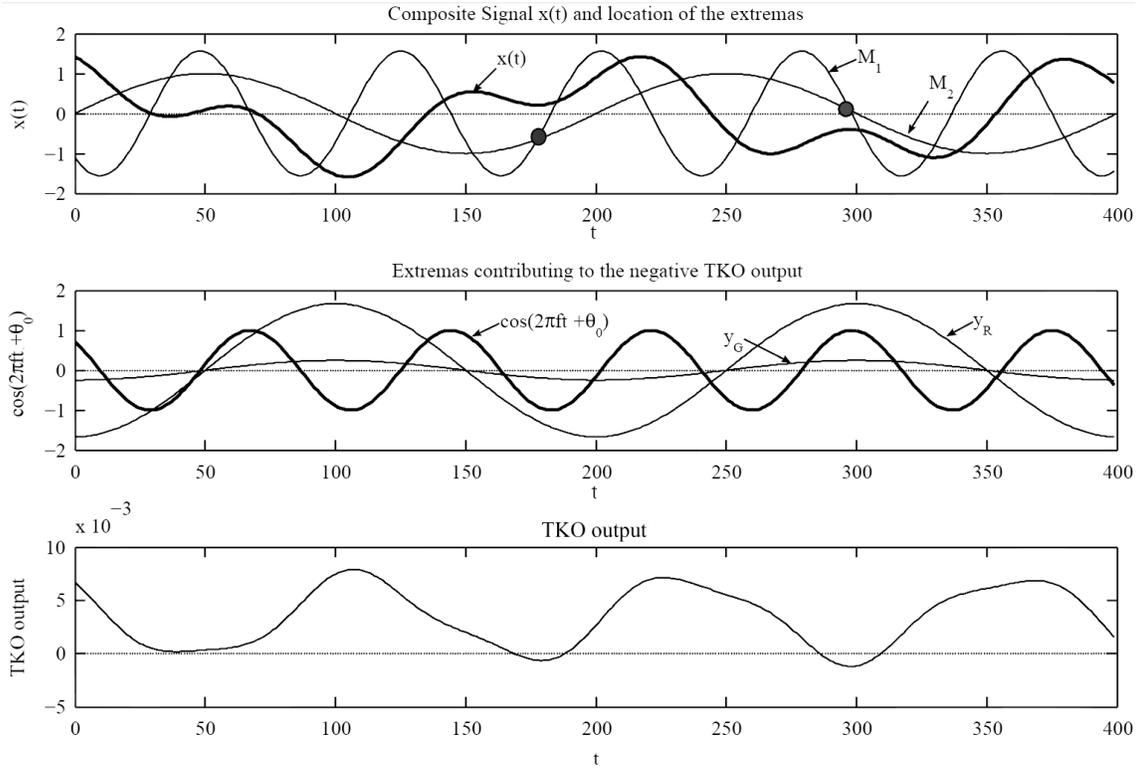}
\caption{TKO Response for two sinusoid input: $a = 0.6, f = 2.3$. $x(t) = \cos(2\pi t) + a\cos(2\pi ft+\theta_0)$. In the top panel, the derivative of the first component ($M_1$) and the opposite of the derivative of the second component ($M_2$) are plotted, so that their intersections correspond to the extremas of $x(t)$. The marked extrema locations contribute to the zero-crossings of the TKO output shown in the bottom panel. The middle panel shows that whenever $\cos(2\pi ft+\theta_0)$ evaluated at the extrema points lies between the curves marked ${y_G}$ and ${y_R}$, the corresponding extrema of $x(t)$ would contribute to the zero crossing of the TKO output}
\label{fig_sinusoid}
\end{figure*}

For certain parameter ranges in the $a$-$f$ plane, the continuous-time TKO $\Psi(x)=\dot{x}^2-x\ddot{x}$ can go negative, a situation that clearly undermines the physical significance of the operator as an energy detector. The situation is best analyzed at the extremas of $x(t)$, where the probability of such an event occurring is the highest. At an extrema, the derivative of (3) is zero, i.e. $\dot x(t;a,f) \propto \sin (2\pi t) + af\sin (2\pi ft + {\theta _0}) = 0$. The conditions for which the operator output is negative at such an extrema is found by evaluating the condition, $ - x\ddot x < 0$, i.e.,
\begin{align*}
{a^2}{f^2}{\cos ^2}(2\pi ft + {\theta _0}) + {\cos ^2}(2\pi t)\\
+ a({f^2} + 1)\cos (2\pi t)\cos (2\pi ft + {\theta _0}) \leqslant 0 \tag{4}
\end{align*}
The required condition is obtained by solving the inequality (4) which is a quadratic in $\cos (2\pi ft + {\theta _0})$, giving the following result: If $x(t;a,f)$ admits an extremum at $t=t_0$, whenever $\cos (2\pi ft + {\theta _0})$ evaluated at such an extrema lies between the curves $y_R=- \cos (2\pi t)/a$ and $y_G=- \cos (2\pi t)/a{f^2}$, $\Psi(x)$ will be negative. Of course, the operator output can be negative at points other than the extrema, albeit with a lesser probability. The situation is depicted in Fig. 1.

The foregoing analysis for the sum of two sinusoids is easily extended to the more general, sinusoid plus noise input case, by considering a spectral distribution of the noise concentrated solely at a frequency $f$, making the noise a sinusoidal Gaussian process with Rayleigh-distributed amplitude [16].
Intuitively, the most vulnerable are the inflection points given by the positive minimas $(x > 0,\;\dot x = 0,\;\ddot x > 0)$ and the negative maximas $(x < 0,\;\dot x = 0,\;\ddot x < 0)$. The zero-crossing rate of the operator output as well as the distribution of the time duration for which the operator output remains negative provide valuable insights into the problem at hand. In particular, the average time duration for which the operator output remains negative can provide valuable insight in deciding on appropriate delay parameters (e.g., $t_p,\;t_q$ for $\Psi_p^q$) or a suitable integration time constant in the post-detection filtering stage.

\section{Statistical Distributions of TKO Outputs\\ and Their Ratios}
\subsection{Probability Density Functions of the TKO outputs}
\emph{Notation:} $\operatorname{tr}(\cdot)$, $(\cdot)'$, $(\cdot)^\dagger$ denote respectively, the trace, nonconjugate and conjugate transpose operators. $\mathbb{E}[\cdot]$ is the expectation operator and $\operatorname{Im}[\cdot]$ is the imaginary part operator. $i=\sqrt{-1}$ is the imaginary unit. The characteristic function (chf) of a random variable (RV) $X$ is $\phi_X(\xi)=\mathbb{E}[\exp{(i\xi X)}]$. We write $f_X(x)=\operatorname{Pr}\{X=x\}$ to denote the probability density function (pdf) and $F_X(x)$ to denote the cumulative distribution function (cdf) of RV $X$. 

We first consider the case when the signals at the detector input are contaminated with additive, zero-mean, stationary Gaussian noise with power spectral density $N_0$.
Let $X$ be the column matrix formed from $r$ real signal plus noise samples $x_j$, $j=1,\ldots,r$.
Let $\mu=\mathbb{E}X$ be the matrix of the means and $M=\mathbb{E}[(X-\mathbb{E}X)(X-\mathbb{E}X)']$ the positive definite (unnormalized) covariance matrix, 
i.e., $X \sim \mathcal{N}_r(\mu,M)$. 
For the jointly Gaussian random variates $(x[n-q],x[n],x[n+q])$, the matrix of the symmetric kernel $\Psi_0^q$ is given by $J=\left(\begin{smallmatrix}0&0&-\frac{1}{2}\\0&1&0\\-\frac{1}{2}&0&0\end{smallmatrix}\right)$. 
Likewise, for jointly distributed RVs $(x[n-q],x[n-p],x[n+p],x[n+q])$, the matrix of the symmetric kernel $\Psi_p^q$ is given by $J=\left(\begin{smallmatrix}0&0&0&-\frac{1}{2}\\0&0&\frac{1}{2}&0\\0&\frac{1}{2}&0&0\\-\frac{1}{2}&0&0&0\end{smallmatrix}\right)$. 

Equivalently, the TKO outputs can be viewed as a linear combination of quadratic detectors with the product correlations reduced to a sum and difference of squares. Thus, for instance, for jointly distributed Gaussian RVs $(x[n],s[n],d[n])$ where $s[n]=(x[n+q]+x[n-q])/2$ and $d[n]=(x[n+q]-x[n+q])/2$, $\Psi_0^q[x[n]]=x^2[n]-s^2[n]+d^2[n]$ and $J=\operatorname{diag}(1,-1,1)$.

Given $J$, the related quadratic form in $X$ is $V(X)= X'JX$.
For a zero-mean, stationary random process whose spectrum is Gaussian, the covariance kernel may take the form $R(t)= \exp{(-ct^2)}$. 
Letting $M=LL'$, $Y=L^{-1}X$ and $Z=Y-L^{-1}\mu$, we have $Z \sim N_r(0,I)$, where $I$ is the identity matrix. 
Let $P$ be a $r\times r$ orthogonal matrix which diagonalizes $L'JL$, so that $P'L'JLP = diag(\lambda_1,\ldots,\lambda_r)=\Lambda$ and $PP'=I$, where $\lambda_1,\ldots,\lambda_r$ are the eigenvalues of $MJ$. Then, letting $U=P'Z$, we have $U \sim \mathcal{N}_r(0,I)$ and $V(X)=(Z+L^{-1}\mu)'L'JL(Z+L^{-1}\mu)=(U+s)'\Lambda (U+s)=\sum\nolimits_{{{j = 1}}}^{{r}} {{\lambda_{{j}}}{{({{{U}}_{{j}}}{{ + }}{{{s}}_{{j}}})}^2}}=\sum\nolimits_{{{j = 1}}}^{{r}} \lambda_j W_j^2$, where $s=P'L^{-1}\mu=(s_1,\ldots,s_r)'$ and ${W_j^2}'$s are independent $\chi^2$ RVs with a single degree of freedom and noncentrality parameter $s_j^2$, i.e., $W_j^2\text{ } \mathop \sim \limits^{{\text{ind}}}\text{ }\chi_1^2 (s_j^2)$. For an indefinite quadratic form where the $\lambda_j'$s can be any real number, letting $\lambda_j>0$ for $j=1,\ldots,k$, $\lambda_j<0$ for $j=k+1,\ldots,l$, and $\lambda_j=0$ for $j=l+1,\ldots,r$, it is easy to see that $V=V_1-V_2$, where 
$V_1=\sum\nolimits_{{{j = 1}}}^{{k}}\lambda_j W_j^2$ and $V_2=\sum\nolimits_{{{j = k+1}}}^{{l}} -\lambda_j W_j^2$. Thus, when $X \sim \mathcal{N}_r(\mu,M)$, $M > 0$, indefinite quadratic forms in $X$
are distributed as the difference of two weighted sums of independent $\chi^2$ RVs [17]. 

The characteristic function (chf) of $V(X)$ is given by 
\begin{align*}
{\phi _V}(\xi ) = \frac{\exp \left\{ { - \tfrac{1}{2}\mu'[I - {{(I - 2i\xi MJ)}^{ - 1}}]{M^{ - 1}}\mu } \right\}}{|I - 2i\xi MJ{|^{\frac{1}{2}}}} \tag{5}
\end{align*}
Alternatively, (5) can be put into the following diagonal form [27], [17], [18], 
\begin{align*}
{\phi_V}(\xi ) = \prod\limits_{j=1}^r {{(1 - 2{{i}}\xi {\lambda _{{k}}}{N_0})}^{ - 1/2}} \exp \{\frac{{{s}}_{{j}}^2}{{2{N_0}}}\frac{{2{{i}}\xi {\lambda _{{j}}}{N_0}}}{1 - 2{{i}}\xi {\lambda _{{j}}}{N_0}}\} \tag{6}  
\end{align*}

For the more specialized case of narrowband inputs, the correlator outputs can be represented as a linear combination of independent, scaled non-central $\chi^2$ variables with two degrees of freedom [18], [19]. 
Now the noise power per cycle $N_0$ is equally split between the in-phase and quadrature components, so that for each in-phase component of noise at the detector input, an additional independent quadrature component of noise is added to the output. 
The latter contributes an extra signal independent term so that in effect, the chf in (6) is multiplied by its value for noise alone and we get [26], [27],
\begin{align*}
{\phi_V}(\xi ) = \prod\limits_{j=1}^r {{(1 - {{i}}\xi {\lambda _{{k}}}{N_0})}^{ - 1}} \exp \{\frac{{s_j}^2}{{{N_0}}}\frac{{{{i}}\xi {\lambda _{{j}}}{N_0}}}{1 - {{i}}\xi {\lambda _{{j}}}{N_0}}\} \tag{7}  
\end{align*}
Formally, let $C$ be the column matrix formed from $r$ complex RVs, $c_j=a_j+ib_j$, $j=1,\ldots,r$, where $a_j$, $b_j$ are normally distributed RVs with 
$\mathbb{E}[(a_k-\mathbb{E}a_k)(a_l-\mathbb{E}a_l)]=\mathbb{E}[(b_k-\mathbb{E}b_k)(b_l-\mathbb{E}b_l)]$, and  $\mathbb{E}[(a_k-\mathbb{E}a_k)(b_l-\mathbb{E}b_l)]=-\mathbb{E}[(a_l-\mathbb{E}a_l)(b_k-\mathbb{E}b_k)]$
Let $\mathbb{E}C=\bar{C}$ and $L=\mathbb{E}[(C-\mathbb{E}C)(C-\mathbb{E}C)^\dagger]$ 
be the complex (nonsingular) covariance matrix. Then, given any Hermitian matrix $Q$, the chf of the (real) quadratic form $V(C)={C}^\dagger Q{C}$ is given by [19]
\begin{align*}
{\phi _V}(\xi ) = \frac{\exp \left\{ { -\bar{C}^\dagger[I - {{(I - i\xi LQ)}^{ - 1}}]{L^{ - 1}}\bar{C}} \right\}}{|I - i\xi LQ|} \tag{8}
\end{align*}
when a straightforward calculation of the matrix operations leads to (7).

In general, inversion of the chf does not yield closed-form solutions for the pdf, except for a few trivial cases. For a single eigenvalue $\lambda$, using the Laplace transform pair 29.3.77 [20], inversion of the chf in (6) yields [27]
\[{f_V}(v) = \frac{{\exp \left\{ { - \tfrac{1}{{2{N_0}}}[{s^2} + \tfrac{v}{\lambda }]} \right\}}}{{\sqrt {2\pi {N_0}\lambda v} }}\cosh \left( {\frac{{s}}{{{N_0}}}\sqrt {\frac{v}{\lambda }} } \right),{\text{ }}v > 0,\]  
and zero otherwise. 
Likewise, for inversion of the chf in (7), the equivalent expression is that of a Rician distribution for the envelope [26].
\[{f_V}(v) = \frac{{\exp \left\{ { - \tfrac{1}{{{N_0}}}[{s^2} + \tfrac{v}{\lambda }]} \right\}}}{{\lambda {N_0}}}{I_0}\left( {\frac{{2|s|}}{{{N_0}}}\sqrt {\frac{v}{\lambda }}}\right),{\text{ }}v > 0,\]   
and zero otherwise, where $I_0(\cdot)$ is the zeroth-order modified Bessel function of the first kind.

For positive only eigenvalues, one can use non-causal convolution of the individual density functions corresponding to the independent eigenmodes yielding the composite density function $f_{V_+}(v_+)$, and similarly for negative only eigenvalues, the density function, $f_{V_-}(v_-)$. The final density function $f_V(v)$ of $V$ is obtained by convolution of the independent densities.
\begin{align*}
{f_V}(v) = \int\limits_{\max (0,v)}^\infty  {d{v_+}f_{V_+}({v_+})f_{V_-}(v - {v_+})} \tag{9}
\end{align*}
If $v > 0$, the lower limit of $v_+$ is $v$, while if $v < 0$, the lower limit is zero. 

As an example, consider the central case ($\mu=0$) with eigenvalues $\lambda_1$, $\lambda_2 >0$ and $\lambda_3$, $\lambda_4 <0$. The corresponding chfs are given by,
$${\phi_V}_+={(1 - 2{{j}}\xi {\lambda _1}{N_0})^{ - 1/2}}{(1 - 2{{j}}\xi {\lambda _2}{N_0})^{ - 1/2}}$$   
$${\phi_V}_-={(1 + 2{{j}}\xi {\lambda _3}{N_0})^{ - 1/2}}{(1 + 2{{j}}\xi {\lambda _4}{N_0})^{ - 1/2}}$$   
Using the Laplace transform pair 29.3.49 [20], the corresponding density functions are
$$f_{V_+}(v_+) = \tfrac{\exp \{  - \tfrac{{v_+}}{{4{N_0}}}(\lambda _1^{ - 1} + \lambda _2^{ - 1})\}{{{I}}_0}(\tfrac{{{v_+}}}{{4{N_0}}}(\lambda _1^{ - 1} - \lambda _2^{ - 1}))}{{2{N_0}\sqrt {{\lambda _1}{\lambda _2}}}}$$  
for $v_+>0$ and zero otherwise.
$$f_{V_-}(v_-) = \tfrac{\exp \{  + \tfrac{{v_-}}{{4{N_0}}}(\lambda _1^{ - 1} + \lambda _2^{ - 1})\}{{{J}}_0}(\tfrac{{v_-}}{{4{N_0}}}(\lambda _1^{ - 1} - \lambda _2^{ - 1}))}{{2{N_0}\sqrt {{\lambda _1}{\lambda _2}}}}$$ 
for $v_-<0$ and zero otherwise. $J_0(\cdot)$ is the zeroth-order Bessel function of the first kind.
The final density function can be found by convolving $f_{V_+}(v_+)$ and $f_{V_-}(v_-)$ using (9). 

Efficient numerical algorithms exist for evaluating the cdf of a quadratic form [21], which mostly rely on Gil-Pelaez's inversion theorem [22]:
\begin{align*}
F_V(v) = \frac{1}{2} - \frac{1}{\pi }\int\limits_0^\infty  {\frac{{\operatorname{Im} [\exp{(-i\xi v)}\phi_V (\xi )]}}{\xi }} d\xi \tag{10}
\end{align*}

The cumulant generating function  admits a power series expansion ${K_V}(\xi ) = \log {\phi _V}(\xi ) = \sum\nolimits_{s = 1}^\infty  {{\kappa _s}} \tfrac{{{{i\xi} ^s}}}
{{s!}}$, and the cumulants $\kappa _s$ are defined by the following identity [23]:
\begin{align*}
{\kappa _s} = {2^{s - 1}}(s - 1)!\{ \operatorname{tr}{(MJ)^s} + s\mu 'J{(MJ)^{s - 1}}\mu \} \tag{11}
\end{align*}
The cumulants $\kappa_1$ and $\kappa_2$ are the mean and the variance, respectively. 
For a pair of independent RVs $(X,Y)$, $\kappa_s(pX+qY)=p^s\kappa_s(X)+q^s\kappa_s(Y)$.
For $s > 2$, the standardized cumulants given by ${\rho _s} = {\kappa _s}\kappa _2^{ - s/2}$, are invariant under affine transformations and provide summary measures of departure from normality in the sense that for a normal distribution, $\kappa_s=0$ for $s > 2$. In particular, $\rho_3$ measures the skewness and $\rho_4$ measures the kurtosis of the distribution.

\subsection{Probability Density Functions of the Ratio of Two TKO Outputs}
The statistical properties of ratios of quadratic forms, their moments, densities and distributions, or approximations thereof, have been extensively studied [23]--[25], [29]. The reader is referred to [17] for a survey of the mainstream statistical literature. 
In general, for any two real functions $V_1$ and $V_2$ (which need not be quadratic forms), the distribution of the ratio, $R=\tfrac{V_1}{V_2}$ can be computed if the joint density function of $V_1$ and $V_2$, $f_{V_1,V_2}(v_1,v_2)$ is known. 
Assuming that $V_2$ is not negative, the probability density of $R$ is given by 
$$f_R(r) = \int\limits_0^\infty  {{v_2}f_{V_1,V_2}(r{v_2},{v_2})d{v_2}}.$$
However, it is very rarely that the joint density is known or $V_1$ and $V_2$ are independent in which case a simple relation applies. When $V_1$ and $V_2$ are not necessarily independent, Geary [28] showed that the density of the ratio $R$ can be written as
\begin{align*}
f_R(r) = \frac{1}{{2\pi i}}{\int\limits_{ - \infty }^\infty  {\left[ {\frac{{\partial \phi_{V_1,V_2} ({\xi _1},{\xi _2})}}{{\partial {\xi _2}}}} \right]} _{{\xi _2} =  - r{\xi _1}}}d{\xi _1} \tag{12}
\end{align*}

When $V_1$ and $V_2$ are noncentral quadratic forms in correlated Gaussian RVs, an efficient numerical algorithm is given in [29] that uses Geary's formulation for computing the density function of the ratio $R$. 
The technique can be used for computing the pdf of the ratios $\frac{\Psi [\dot x(t)]}{\Psi [x(t)]}$ and $\frac{(\Psi [x(t)])^2}{\Psi [\dot x(t)]}$ for the instantaneous frequency squared and instantaneous amplitude or envelope squared estimates, respectively. For the latter, the density function for the numerator which is of the form, $V_3=V_1^2$ can be obtained from the density function of $V_1$ by a nonlinear transformation. Correspondingly, $f_{V_3V_2}(v_3,v_2)$ needs be transformed for use in (12), to get the pdf of the ratio $V_3/V_2$. 

\section{Results and Discussion}
An important consequence of the assumption of a stationary Gaussian noise structure at the detector input is that all higher order probability densities are completely specified if the covariance function of the input noise is known. Because of stationarity, the covariance matrix has elements $m_{ij}=\mathbb{E}[x_ix_j]$, which depend only on the absolute difference or distance $\lvert i-j \rvert$ between the samples and accordingly, a single subscript $m_j=\mathbb{E}[x_ix_j]$, $i, j = 0,\pm 1,\pm 2,\ldots$ etc., suffices, where $x_i \equiv x[i]$. The covariance function of the input noise process is taken to be a Gaussian kernel given by $\exp(-t^2/2)$. Smooth estimates of the signal derivative are obtained from a finer grid by interpolating the signal samples. The first few moments of the TKO output are easy to find. For instance, writing the output of $\Psi_p^q$ as $(s_1+n_1)(s_2+n_2)-(s_3+n_3)(s_4+n_4)$, where $s_i$ and $n_i$  are the signal and noise components at sample point $i$, we have $\mathbb{E}[\Psi_p^q]=(s_1s_2-s_3s_4)+(m_{2p}-m_{2q})$. The variance $\operatorname{Var}[\Psi_p^q]$ has the form ${N_0}\sum\nolimits_{i = 1}^4 {s_i^2}  + 2[{s_1}{s_2}{m_{2p}} +$ ${s_3}{s_4}{m_{2q}} - ({s_1}{s_3} + {s_2}{s_4}){m_{q - p}} - ({s_1}{s_4} + {s_2}{s_3}){m_{q + p}}]$. We see that there is an average term contributed solely by noise, as well as a few signal$\times$noise terms that cause strong time-varying fluctuations. For a time-varying signal, the mean, variance, and the covariance are also time-varying and an averaging over one complete cycle of the input sinusoid is necessary to have meaningful expressions for the mean, variance as well as the higher-order moments. Also, signal and signal derivative product terms lead to the creation of second harmonics, and a suitable post-filtering action is necessary. 
\begin{figure}[!t]
\centering
\includegraphics[width=3.5in]{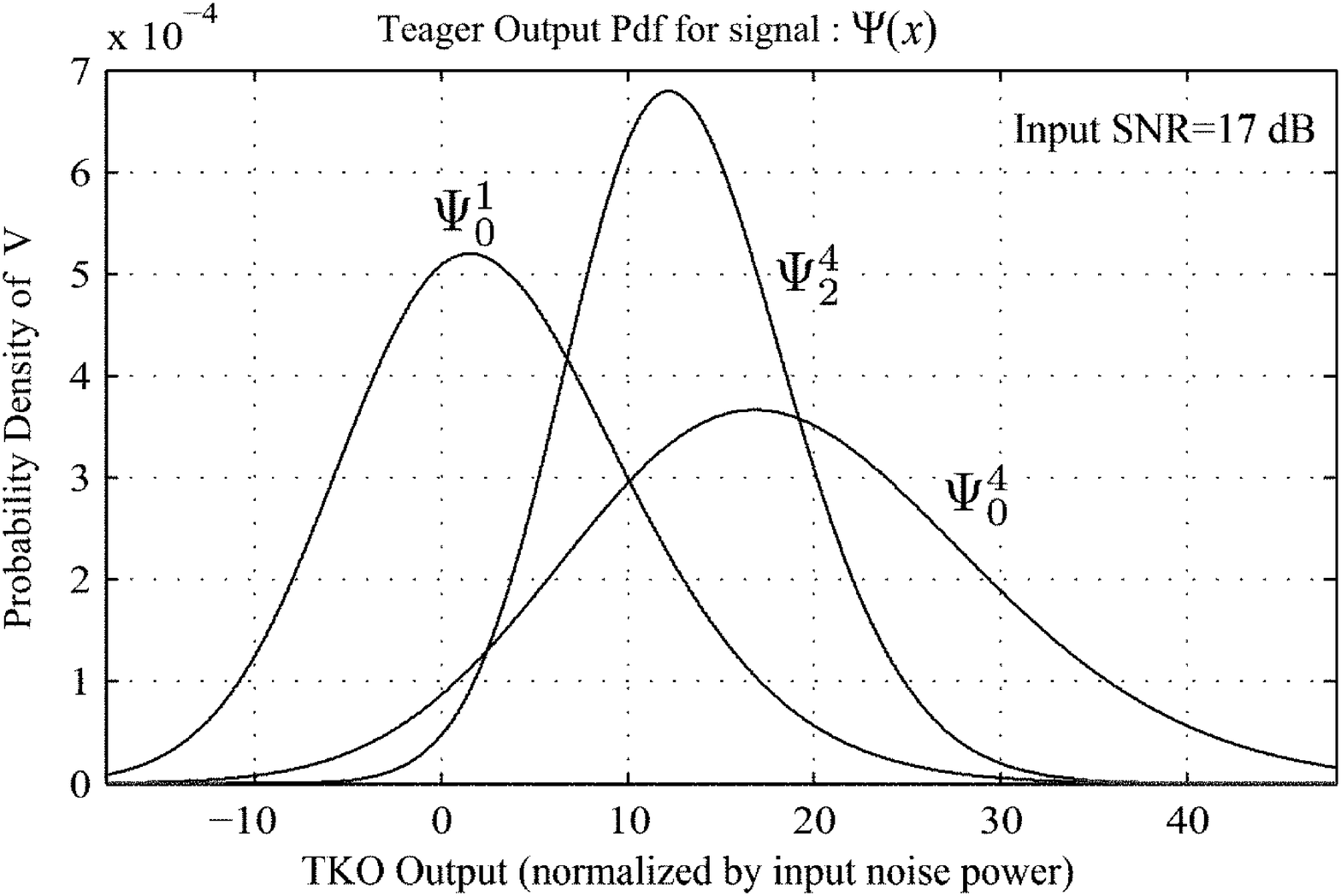}
\includegraphics[width=3.5in]{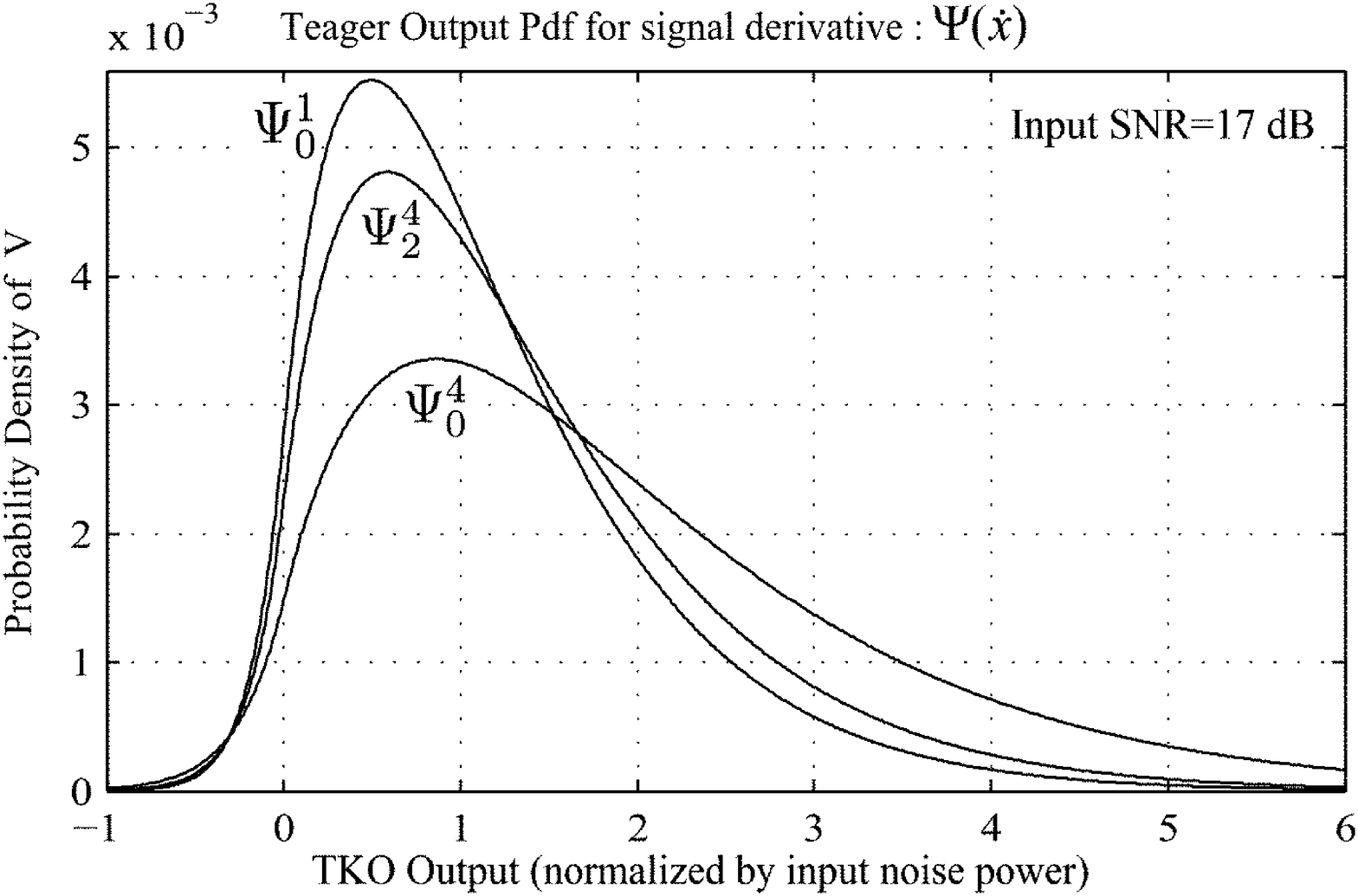}
\includegraphics[width=3.55in]{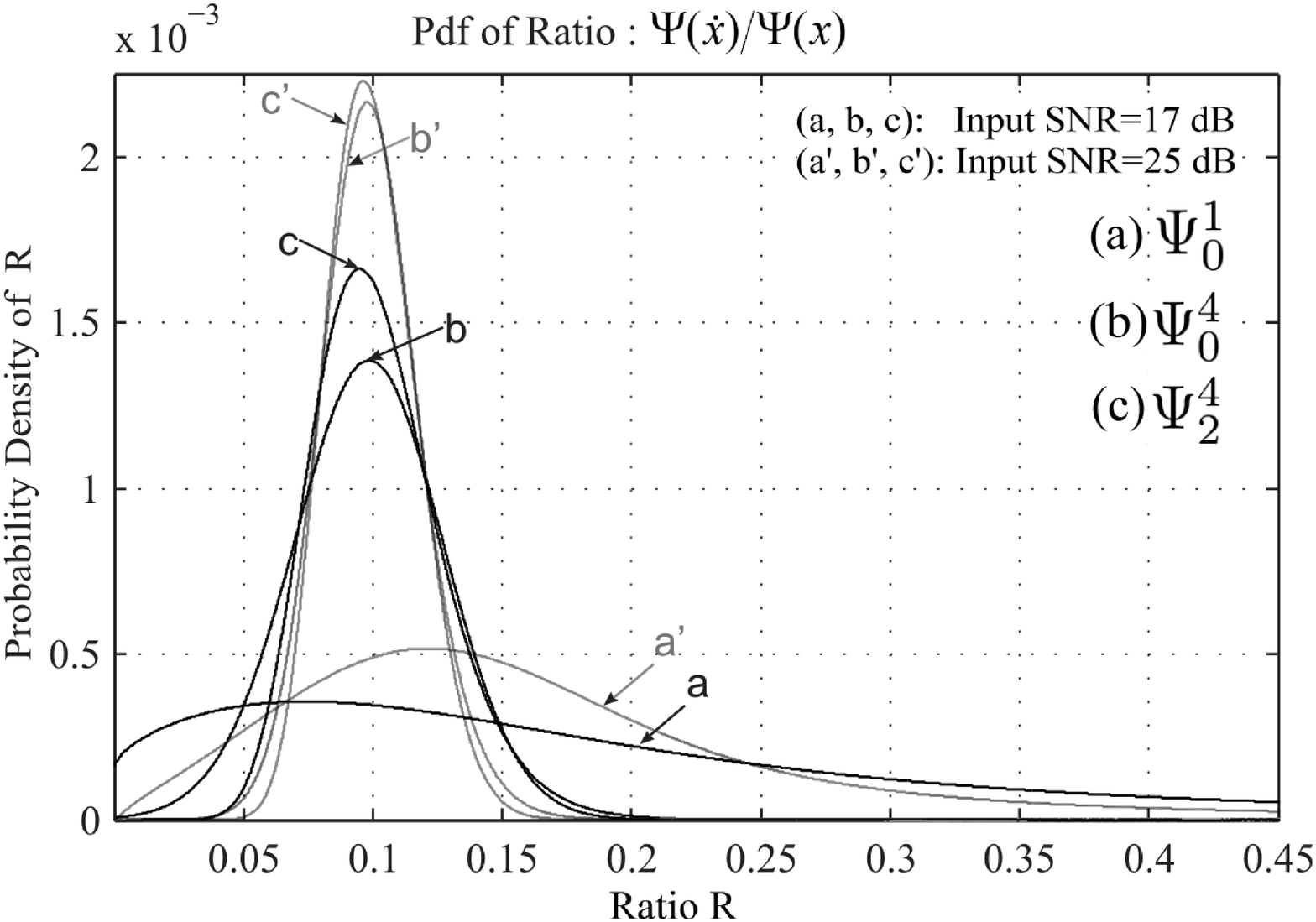}
\caption{Pdfs for the TKO outputs (normalized by the noise power into detector) for the signal, signal derivative and the ratios (Input SNR = 17 dB). For the distribution of the ratios, two different sets of pdfs show the effect of noise power into detector: the sets (a,b,c) and (a',b',c') correspond to an input SNR of 17 dB and 25 dB respectively}
\label{l1}
\end{figure}

The computed probability densities of the outputs of different TKO variants, as well as the ratio $\frac{\Psi [\dot x(t)]}{\Psi [x(t)]}$ which yields the instantaneous frequency squared are shown in Fig. 2 for different input signal-to-noise ratios (SNRs).
As expected, the mean output SNR $\mathbb{E}[\kappa_{1(S+N)}/\kappa_{1(N)}]$ improved as the delay parameters ($t_p,t_q$) are increased. 
For the TKO applied to the signal, at an input SNR of 17 dB, the output SNRs for the operators $\Psi_0^1$, $\Psi_0^4$ and $\Psi_2^4$ are respectively (a) 6.65 dB, (b) 24.92 dB, and (c) 29.52 dB. 
Evidently, the four-sample input TKO $\Psi_p^q$ provides a better noise performance compared to a three-sample input TKO $\Psi_0^q$. This can be attributed to the fact that the mean and variance of the noise terms are larger in the latter case. Similar remarks hold for the TKO applied to the signal derivative. 

Furthermore, depending upon the structure of the operator kernel $J$ and the input noise correlation, a finite number of eigenvalues associated with the product $MJ$ in (5) are always found to be negative.
For instance, at an input SNR of 17 dB, the eigenvalues of $MJ$ computed for the operators
$\Psi_0^1$, $\Psi_0^4$ and $\Psi_2^4$ are respectively, $(.836, .475, -.361)$, $(.952,.498,-.455)$, and $(.655,.376,-.321,-.042)$. Negative eigenvalues correspond physically to the fact that the operator output $V(t)$ can go negative for some time $t$ ($> 0$), although there can still be a positive mean value.
If the delay parameters ($t_p, t_q$) are large (vis-\`a-vis the correlation time of the input noise, $\delta_c$), we have a large number of eigenvalues and an asymptotically normal density function can be expected.
Conversely, for small delay parameters we expect large departures from normality, as is evidenced in the plots in Fig. 2. 
From Fig. 2, it is also evident that the probability of excursion towards the negative region is lower for $\Psi_2^4$ compared to $\Psi_0^4$ and $\Psi_0^1$. 

\begin{figure}[!t]
\centering
\includegraphics[width=3.5in]{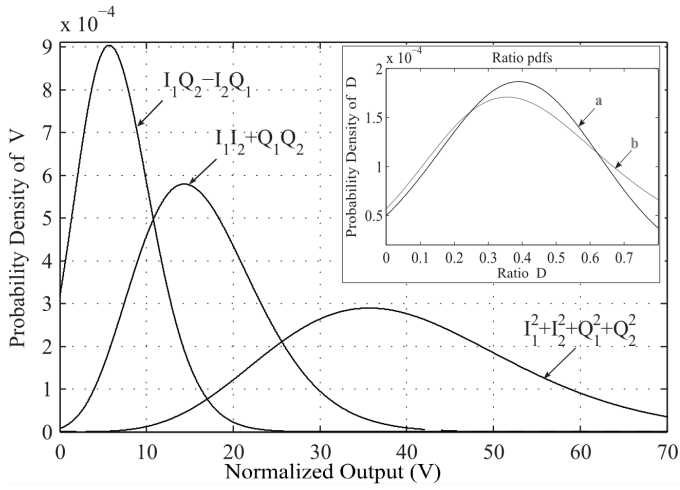}
\caption{Pdfs for the individual quadratic forms: $(I_1Q_2-I_2Q_1)$, $(I_1Q_2 + I_2Q_1)$, and $(I_1^2+I_2^2+Q_1^2+Q_2^2)$. Input SNR = 9.04 dB. \emph{Inset}: pdfs of the sine and tangent I-Q correlators (a) $2(I_1Q_2-I_2Q_1)/(I_1^2+I_2^2+Q_1^2+Q_2^2)$ and (b) $(I_1Q_2 - I_2Q_1) / (I_1Q_2 + I_2Q_1)$}
\label{l2}
\end{figure}
For narrowband inputs, ratios of quadratic forms that yield the sine and tangent of the phase difference are respectively, (a) $2(I_1Q_2-I_2Q_1)/(I_1^2+I_2^2+Q_1^2+Q_2^2)$ and (b) $(I_1Q_2 - I_2Q_1) / (I_1Q_2 + I_2Q_1)$, where $I_i$ and $Q_i$ are the in-phase and quadrature components, $i=1, 2$ denoting two adjacent samples. Fig. 3 shows the density functions of the individual quadratic forms and their ratios. For the quadratic form $(I_1Q_2-I_2Q_1)$, the mean and the higher odd order moments of the noise term are zero. It is easy to show that the covariance of $(I_1Q_2-I_2Q_1)$ and $(I_1^2+I_2^2+Q_1^2+Q_2^2)$ is greater than that of $(I_1Q_2 - I_2Q_1)$ and $(I_1Q_2 + I_2Q_1)$. This gives a marginally higher variance for the ratio in (b) compared to that in (a). Pdf of the ratios shown in the inset of Fig. 3 confirms these observations.
\begin{figure}[!t]
\centering
\includegraphics[width=3.0in]{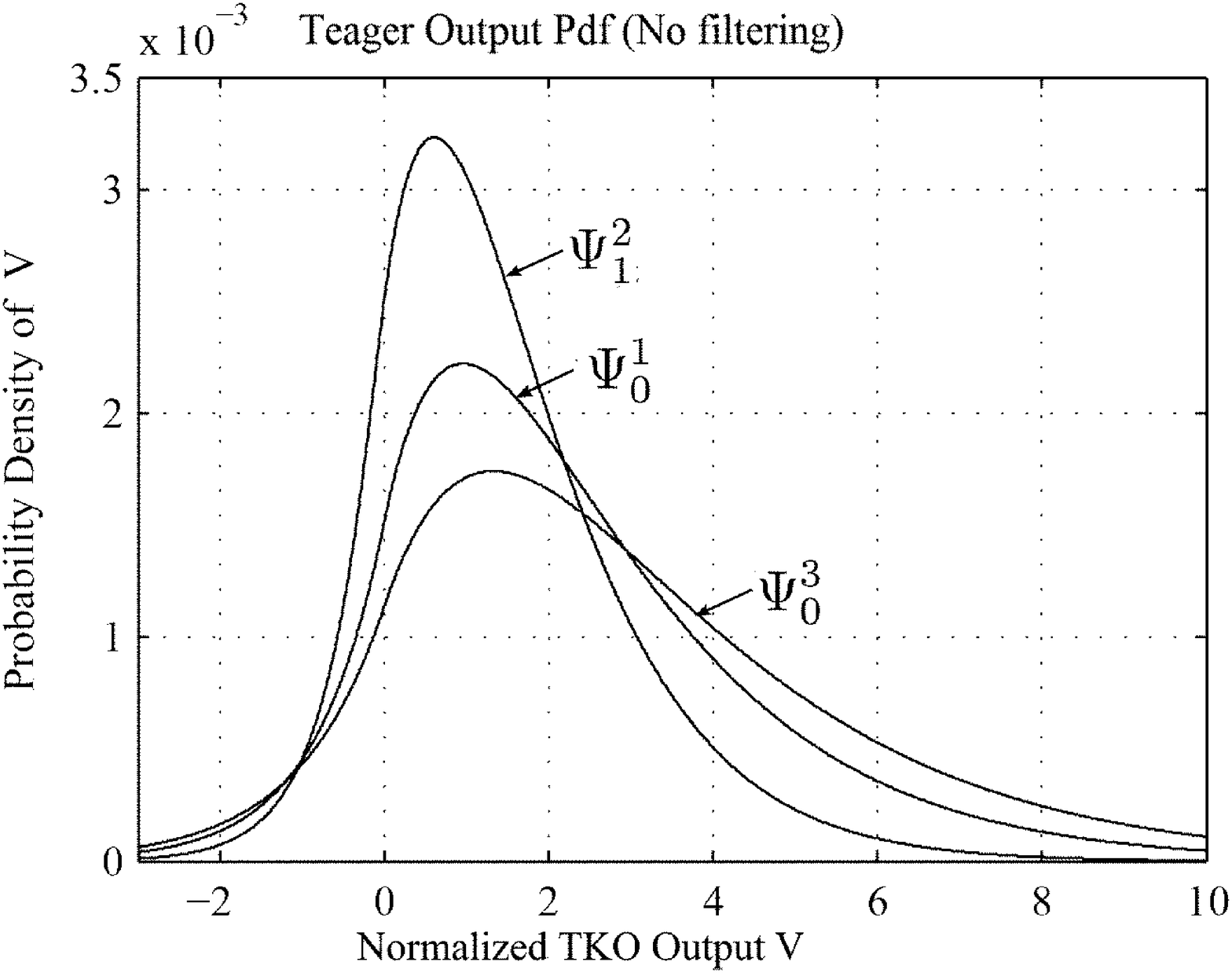}
\includegraphics[width=3.0in]{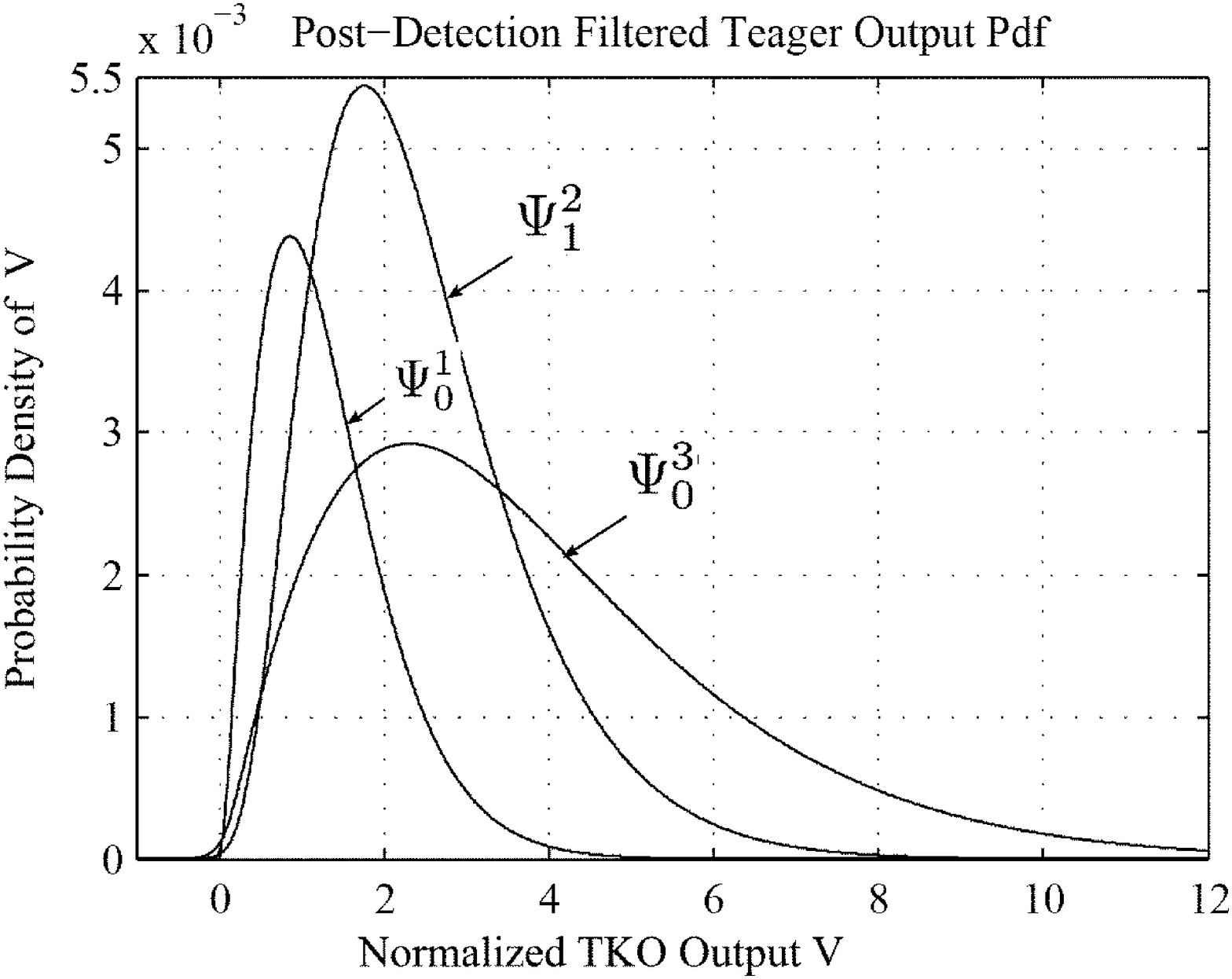}
\caption{TKO output before and after post-detection filtering using a three-point Binomial filter with an impulse response $(1, 2, 1)$. Input SNR = 11 dB.}
\label{l4}
\end{figure}

A way of increasing the signal mean value and ensuring the positivity of the operator outputs is to sum several TKO outputs. 
If the observation time is long enough to allow for decorrelation in time, the constituent outputs are independent and the cumulants add.
This ensures an increase in the mean value. 
Positivity can also be achieved by introducing a spectrally narrow low-pass filter at the TKO post-detection stage (e.g., a filter with a long time constant compared to $\delta_c$). 
Integration attenuates high frequency spectral components and is an effective option for filtering the strong time-varying fluctuations induced by the signal$\times$noise terms.
A long time constant however compromises the ``instantaneous'' property of the TKO. Results of applying a three-point binomial filter at the TKO output $\Psi [\dot x(t)]$ is shown in Fig. 4. 
The short impulse response $(1, 2, 1)$ of the filter preserves the superior localization property of the TKO and restores positivity of the operator outputs.
When filtering is not an option, simple thresholding to remove negative outputs or outputs near zero is a viable alternative that can retain the speed advantage of the operator, if the penalty in terms of missed data is not a concern.

\vspace{2mm}
\section*{Acknowledgments}
\vspace{2mm}
Thanks are due to Prof. Tarun K. Bhattacharyya for useful discussions. A part of this work was funded by the Advanced VLSI Consortium, IIT Kharagpur.

\begin{IEEEbiography}[{\includegraphics[width=1in,height=1.25in,clip,keepaspectratio]{picture}}]{John Doe}
\blindtext
\end{IEEEbiography}


\begin{thebibliography}{99}
\vspace{1.1mm}
\balance
\bibitem{l2} H. M. Teager and S. M. Teager, ``Evidence for nonlinear sound production mechanisms in the vocal tract,'' in {\em Speech Production and Speech Modeling}, W. J. Hardcastle and A. Marchal, Eds. Norwell, MA: Kluwer, 1990, pp. 241-261.
\vspace{-1.1mm}
\bibitem{l3} J. F. Kaiser, ``On a simple algorithm to calculate the 'energy' of a signal,'' {\em Proc. IEEE Proc. ICASSP '90}, Albuquerque, Apr. 1990.
\vspace{-1.1mm}
\bibitem{l4} P. Maragos, J. F. Kaiser, and T. F. Quatieri, ``On amplitude and frequency demodulation using energy operators,'' {\em IEEE Trans. Signal Process.}, vol. 41, no. 4, pp. 1532-1550, Apr. 1993.
\vspace{-1.1mm}
\bibitem{l5} P. Maragos, J. F. Kaiser, and T. F. Quatieri, ``Energy separation in signal modulations with application to speech analysis,'' {\em IEEE Trans. Signal Process.}, vol. 41, no. 10, pp. 3024-3051, Oct. 1993.
\vspace{-1.1mm}
\bibitem{l6} A. C. Bovik, P. Maragos, and T. F. Quatieri, ``AM-FM energy detection and separation in noise using multiband energy operators,'' {\em IEEE Transactions on Sigal Process.}, vol. 41, pp. 3245-3265, Dec. 1993.
\vspace{-4mm}
\bibitem{l7} Alan C. Bovik and Petros Maragos, ``Conditions for positivity of an energy operator,'' {\em IEEE Trans. Signal Process.}, vol. 42, pp. 469-471, Feb. 1994.
\vspace{-1.1mm}
\bibitem{l8} B. Santhanam and P. Maragos, ``Multicomponent AM-FM demodulation via periodicity-based algebraic separation and energy-based demodulation,'' {\em IEEE Trans. Commun.}, vol. 48, no. 3, pp. 473-490, Mar. 2000.
\vspace{-4mm}
\bibitem{l9} P. Maragos and A. C. Bovik, ``Image demodulation using multidimensional energy separation,'' {\em J. Opt. Soc. Amer.}, vol. 12, no. 9, pp. 1867-1876, 1995.
\vspace{-1.1mm}
\bibitem{l10} W. Lin, C. Hamilton, and P. Chitrapu, ``A generalization to the Teager-Kaiser energy function and application to resolving two closely-spaced tones,'' {\em Proc. IEEE ICASSP-95}, pp. 1637-40, May 1995.
\vspace{-1.1mm}
\bibitem{l11} R. Hamila, E. S. Lohan, and M. Renfors, ``Subchip multipath delay estimation for Downlink WCDMA system based on Teager-Kaiser operator,'' {\em IEEE Commun. Lett.}, vol. 7, no.1, Jan, 2003.
\vspace{-1.1mm}
\bibitem{l12} P. Maragos, A. Potamianos, ``Higher order differential energy operators,'' {\em IEEE Signal Proc. Lett.}, vol. 2, pp. 152-154, Aug. 1995.
\vspace{-1.1mm}
\bibitem{l13} G. W. Groves, ``Probability distribution of complex Monopulse Ratio with arbitrary correlation between channels,'' {\em IEEE Trans. Aero. and Electronic Sys.}, vol. 33, pp. 1345-50, Oct. 1997.
\vspace{-1.1mm}
\bibitem{l14} D. Dimitriadis and P. Maragos, ``An improved energy demodulation algorithm using splines,'' in {\em 2001 IEEE International Conf. Acoustics, Speech, and Signal Process. Proc.}, vol. 6, pp. 3481-4, 2001.
\vspace{-1.1mm}
\bibitem{l15} D. Dimitriadis, A. Potamianos, and P. Maragos, ``A comparison of the squared energy and Teager-Kaiser energy operators for short-term energy estimation in additive noise,'' {\em IEEE Trans. Signal Process.}, vol. 57, pp. 2569-81, July, 2009. 
\vspace{-1.1mm}
\bibitem{l16} J. Fang and L. E. Atlas, ``Quadratic detectors for energy estimation,'' {\em IEEE Trans. Signal Process.}, vol. 43, no. 11, pp. 2582-2594, 1995.
\vspace{-1.1mm}
\bibitem{l17} N. M. Blachman, ``Zero-crossing rate for the sum of two sinusoids or a signal plus noise (Corresp.),'' {\em IEEE Trans. Inf. Theory}, vol. 21, no. 6, pp. 671-675, 1975.
\vspace{-1.1mm}
\bibitem{l18} A. M. Mathai and S. B. Provost, {\em Quadratic Forms in Random Variables}, New York: Marcel Dekker, 1992.
\vspace{-1.1mm}
\bibitem{l19} D. R. Middleton,  {\em Introduction to statistical communication theory}, New York: McGraw-Hill, 1960.
\vspace{-1.1mm}
\bibitem{l20} G. L. Turin, ``The characteristic function of Hermitian quadratic forms in complex normal variables,'' {\em Biometrika}, vol. 47, pp. 199-201, 1960.
\vspace{-1.1mm}
\bibitem{l21} M. Abramowitz and I. A. Stegun, {\em Handbook of Mathematical Functions}, Natl. Bur. Stand., Appl. Math. Series No. 55, Washington, DC, 1964.
\vspace{-1.1mm}
\bibitem{l22} J. P. Imhof, ``Computing the distribution of quadratic forms in normal variables,'' {\em Biometrika}, 48, 419-426, 1961.
\vspace{-1.1mm}
\bibitem{l23} J. Gil-Pelaez, ``Note on the Inversion theorem,'' {\em Biometrika}, 38, 481-2, 1951.
\vspace{-1.1mm}
\bibitem{l24} J. R. Magnus, ``The exact moments of a ratio of Quadratic forms in normal variables,'' {\em Annales d'Économie et de Statistique}, No. 4 (Oct.-Dec., 1986), pp. 95-109.
\vspace{-1.1mm}
\bibitem{l25} R. Lugannini, and S. O. Rice, ``Distribution of the ratio of quadratic forms in normal variables-Numerical methods,'' {\em SIAM. J. Sci. Stat. Comput.}, Vol. 5, pp. 476-88, June 1984.
\vspace{-1.1mm}
\bibitem{l26} T. Y. Al-Naffouri, B. Hassibi, ``On the distribution of Indefinite Quadratic forms in Gaussian random variables,'' in {\em Proc. IEEE ISIT}, Jun. 2009.
\vspace{-1.1mm}
\bibitem{l261} M. Kac, and A. J. F. Siegert, ``On the theory of noise in radio receivers with square-law detectors,'' {\em J. Appl. Phys.}, vol. 18, pp. 383-97, 1947.
\vspace{-1.1mm}
\bibitem{l260} R. C. Emerson, ``First probability densities for receivers with square law detectors,'' {\em J. Appl. Phys.}, vol. 24, pp. 1168-76, 1953.
\vspace{-1.1mm}
\bibitem{l27} R. C. Geary, ``Extension of a theorem by Harald Cramer on the frequency distribution of the quotient of two random variables,'' {\em J. Roy. Statist. Soc.}, 107, pp. 56-57, 1944.
\vspace{-.7mm}
\bibitem{l28} S. Broda, and M.S. Paolella, ``Evaluating the density of ratios of noncentral Quadratic forms in normal variables,'' {\em Comput. Stat. Data Anal.}, 53, pp. 1264-1270, 2009.
\end{thebibliography}
\end{document}